%% file: molendi.tex
\def \SAIT #1 #2 {{\em Mem.\ Soc.\ Astron.\ It.\/} {\bf #1}, #2}
\def \MESS #1 #2 {{\em The Messenger\/} {\bf #1}, #2}
\def \ASTRNACH #1 #2 {{\em Astron. Nach.\/} {\bf #1}, #2}
\def \AAP #1 #2 {{\em Astron. Astrophys.\/} {\bf #1}, #2}
\def \AAL #1 #2 {{\em Astron. Astrophys. Lett.\/} {\bf #1}, L#2}
\def \AAR #1 #2 {{\em Astron. Astrophys. Rev.\/} {\bf #1}, #2}
\def \AAS #1 #2 {{\em Astron. Astrophys. Suppl. Ser.\/} {\bf #1}, #2}
\def \AJ #1 #2 {{\em Astron. J.\/} {\bf #1}, #2}
\def \ANNREV #1 #2 {{\em Ann. Rev. Astron. Astrophys.\/} {\bf #1}, #2}
\def \APJ #1 #2 {{\em Astrophys. J.\/} {\bf #1}, #2}
\def \APJL #1 #2 {{\em Astrophys. J. Lett.\/} {\bf #1}, L#2}
\def \APJS #1 #2 {{\em Astrophys. J. Suppl.\/} {\bf #1}, #2}
\def \APSS #1 #2 {{\em Astrophys. Space Sci.\/} {\bf #1}, #2}
\def \ASR #1 #2 {{\em Adv. Space Res.\/} {\bf #1}, #2}
\def \BAIC #1 #2 {{\em Bull. Astron. Inst. Czechosl.\/} {\bf #1}, #2}
\def \JSQRT #1 #2 {{\em J. Quant. Spectrosc. Radiat. Transfer\/} {\bf #1}, #2}
\def \MN #1 #2 {{\em Mon. Not. R. Astr. Soc.\/} {\bf #1}, #2}
\def \MEM #1 #2 {{\em Mem. R. Astr. Soc.\/} {\bf #1}, #2}
\def \PLR #1 #2 {{\em Phys. Lett. Rev.\/} {\bf #1}, #2}
\def \PASJ #1 #2 {{\em Publ. Astron. Soc. Japan\/} {\bf #1}, #2}
\def \PASP #1 #2 {{\em Publ. Astr. Soc. Pacific\/} {\bf #1}, #2}
\def \NAT #1 #2 {{\em Nature\/} {\bf #1}, #2}

\documentstyle[twoside]{memsait}
\input epsf.sty
\input newpsfig.tex
\begin{opening}
\title{First Results from the MECS on board BeppoSAX}
\author{
S.Molendi$^{1,2}$,
L.Chiappetti$^1$,
G.Cusumano$^3$,
D.Dal Fiume$^4$,
F.Fiore$^2$,
F.Frontera$^4$,
P.Giommi$^2$,
M.Guainazzi$^2$,
C.Maccarone$^3$,
A.Matteuzzi$^2$,
T.Mineo$^3$,
G.C.Perola$^5$,
L.Piro$^6$,
D.Ricci$^2$ and
B.Sacco$^3$
}
\institute{$^1$Istituto di Fisica Cosmica e Tecnologie Relative, Milano, Italy\\
$^2$BeppoSAX Science Data Center, Rome, Italy\\
$^3$Istituto di Fisica Cosmica ed Applicazioni all'Informatica,Palermo,Italy\\
$^4$Istituto Tecnologie e Studio delle Radiazioni Extraterrestri, Bologna, Italy\\
$^5$Dipartimento di Fisica, Universit\`a degli Studi "Roma 3", Rome, Italy\\
$^6$Istituto di Astrofisica Spaziale, Frascati, Italy
}
\date{} 
\end{opening}

\begin{document}

\oddpagefooter{}{}{} 
\evenpagefooter{}{}{} 
\ 
\bigskip

\begin{abstract}
In this contribution we discuss briefly a few calibration items relevant to
the data analysis and present some preliminary scientific results.
The discussion on instrumental topics focuses on the response matrix and 
Point Spread Function (PSF). In the scientific results section we discuss 
a first analysis of the two Seyferts MCG 6-30-15 and NGC 4151 and of 
the Cosmic X-ray Background.   
\end{abstract}

\section{Introduction}
During the Science Verification Phase (SVP) a number of bright X-ray sources
have been observed by BeppoSAX. The aim of such observations is 
twofold: perform inflight calibrations of the instruments and verify the 
capabilities of BeppoSAX to achieve the scientific goals for which it has been
designed.
In this presentation we concentrate on the analysis of SVP data  
from the Medium Energy Concentrator Spectrometer (MECS) on board BeppoSAX.
A more general presentation is given by Piro et al. in these proceedings.

\section{Calibration and Instrumental Issues}
An extensive description of the MECS ground calibrations can be found in 
Boella et al. (1997). In this short presentation we focus onto two of 
the most important calibration aspects that have been addressed using 
inflight data.

\begin{figure}
\centerline{
\hbox{
\psfig{figure=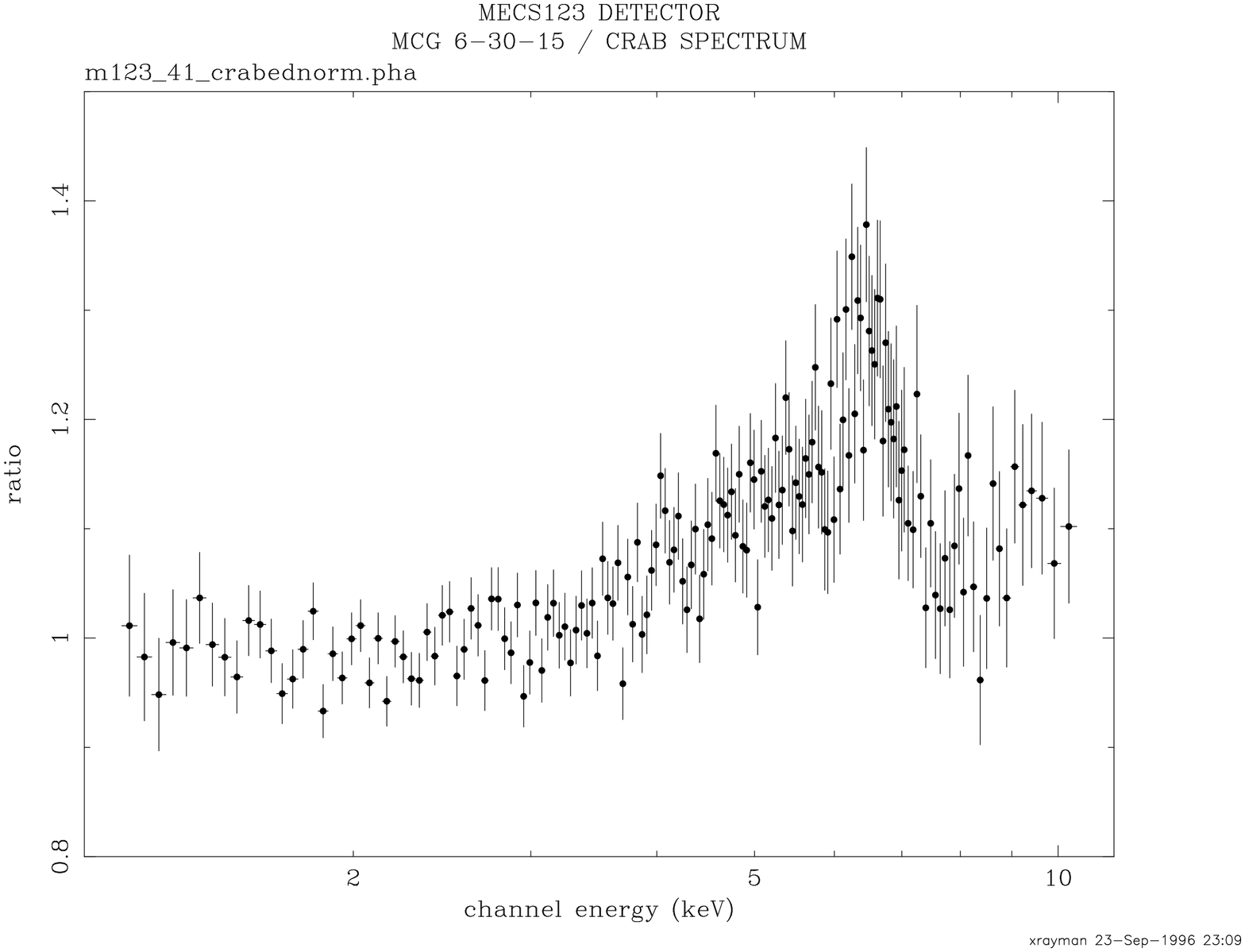,height=5.5truecm,width=8.8truecm,bbllx=28.5mm,bblly=13mm,bburx=201mm,bbury=255mm,angle=-90,clip=}
}
}
\caption{Ratio, in arbitrary units, of the MECS MCG 6-30-15 to Crab Spectrum.}
\end{figure}

\subsection{Response Matrix}
The quality of the ground derived response matrix has been tested
by observing well known bright sources, the most important being the
Crab. For this source 
residuals to a power-law plus absorption fit are contained within 3\% 
everywhere except 
in the two critical regions of the Gold M-edge at 2.2 keV and of the 
Xenon L-edge at 4.7  keV, where they can be as high as $\simeq$ 10\%. 
The measured absolute flux 
in the 2-10 keV band is within 3\% of the nominal value. This is certainly a 
very encouraging result for a ground response matrix. The MECS hardware group 
in Palermo has recently finished working on an inflight matrix
which has now been released to the scientific community. Fitting 
of Crab data with this matrix yields residuals contained virtually 
everywhere within 2\%. A plot of these residuals, as well as other relevant 
information, can be found at the following hyperlink 
{\it http://www.sdc.asi.it/software/cookbook/matrices.html}.

\subsection{Point Spread Function}
The Point Spread Function (PSF) of the MECS results from the convolution
of the telescope PSF and the detector PSF. An energy averaged plot of the SAX
MECS PSF can be found at the following hyperlink: 
{\it http://www.sdc.asi.it/software/cookbook/psf.html}.
The measured on-axis PSF is sensibly smaller than the ASCA-GIS on-axis PSF,
the half power radius at 6 keV being about a factor 2 smaller. 
The comparison is even more favorable at the 80\% and 90\% radii,
due to the considerably reduced scattering of the MECS optics. 
Another important advantage of the MECS-PSF is the moderate off-axis
degradation. 
 
\section{First Scientific Results}
\subsection{MCG 6-30-15}
During the SVP phase we have carried out a 130 ks observation of 
the Seyfert galaxy MCG 6-30-15. The MECS lightcurve shows strong (more 
than a factor 2) and rapid (halfing timescale $<$ 2000 s) variability,
consistently with the ROSAT (Nandra \& Pounds 1992) and ASCA 
(Otani et al. 1996) 
observation of this source.
An ASCA observation of this source (Tanaka et al. 1995) has evidenced
an excess emission with respect to a power-law model in the energy 
range 4-6 keV. This feature has been interpreted as the red tail of the Iron
K$_\alpha$ line peaking around 6.4 keV. Such a tail results from the 
combination 
of  Doppler and gravitational redshifts when the emitting region is located 
in the immediate vicinity of a massive blackhole.
As pointed out by Tanaka and collaborators, the detection 
of the extended red tail of the Fe K$_\alpha$ is of great importance since it
can be considered one of the most direct manifestations of the 
blackhole at the center of active galaxies. 

To investigate the presence of the excess in the 4-6 keV band discovered
by ASCA we have divided the MCG-6-30-15 spectrum 
by the Crab Spectrum.
In such a way any residual calibration systematics at the Xenon
edge energy ($E \simeq 4.7$keV) does not affect our result.
In order to increase the statistics, data from all 3 MECS 
has been used.
As can be seen in Figure 1 our observation confirms the presence of the 
excess in the 4-6 keV band.
We consider this result of some importance as it is the first confirmation
of such a spectral feature from a satellite other than ASCA.

Having confirmed the presence of the feature, clearly the most important
thing to do is to verify its proposed origin (i.e. red tail of the K$_\alpha$
line)
by making use of information that can be extracted only from the BeppoSAX
observation. One of the key aspects in 
understanding the nature of the excess in the 4-6 keV band is a reliable 
estimate of the underlying continuum. In this band a contribution to 
the emission comes from the reflection component observed by GINGA in
this and other Seyferts (Nandra \& Pounds 1994). The entity of this 
contribution could not be easily assessed with ASCA as the effective areas 
of the mirrors drop rapidly 
above 7 keV where the contribution from the reflection component rises 
sharply. The BeppoSAX observatory, with its high energy instruments,
can perform measurements around 40 keV where the reflection component
peaks. 
\begin{figure}
\centerline{
\hbox{
\psfig{figure=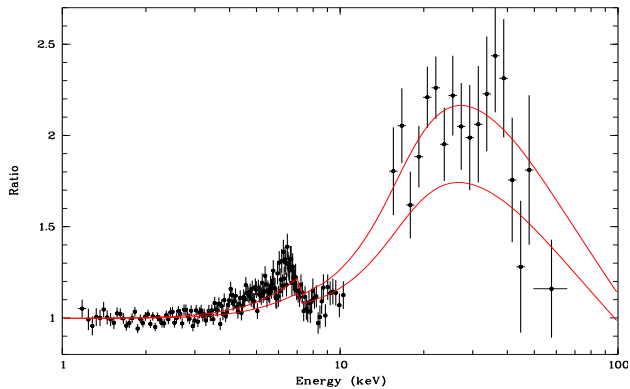,height=5.5truecm,width=8.8truecm,bbllx=8.5mm,bblly=13mm,bburx=201mm,bbury=275mm,angle=-90,clip=}
}
}
\caption{Ratio, in arbitrary units, of the MECS $+$ PDS MCG 6-30-15 to Crab Spectrum.
The two lines represent power-law + reflection models with normalization of
the reflection component ($\Omega /2 \pi $) set respectively to 1 (lower line)
and 1.5 (upper line).}
\end{figure}
In Figure 2 we show the combined MECS PDS spectrum of MCG 6-30-15.
To avoid intercalibration problems between the 2 instruments we use again
the ratio of the MCG 6-30-15 to the Crab spectrum. The lower line indicates 
a powerlaw plus reflection model with the normalization of the reflection 
component ($\Omega/ 2\pi$) set to 1 as in the fit reported by Tanaka and
collaborators on the ASCA data. The model clearly falls short of the
data. A reasonable agreement between the two can be achieved by increasing the
normalization of the reflection component by about 50\% (higher line). 
If the extrapolation of the reflection component used in the ASCA fit 
underestimates the data at 40 keV it may be
reasonable to assume that the model applied to fit the ASCA data also
underestimates the contribution of this component in the 4-6 keV band. 
Following this line of reasoning, if we use a model with reflection
normalization $\Omega/ 2\pi = 1.5$ we find that: 1) a large fraction of 
the flux previously 
attributed to the red wing is now produced by the reflection component 
2) the data is still in excess of the model, suggesting the presence of an
emission component other than the powerlaw and the reflection.
\begin{figure}
\centerline{
\hbox{
\psfig{figure=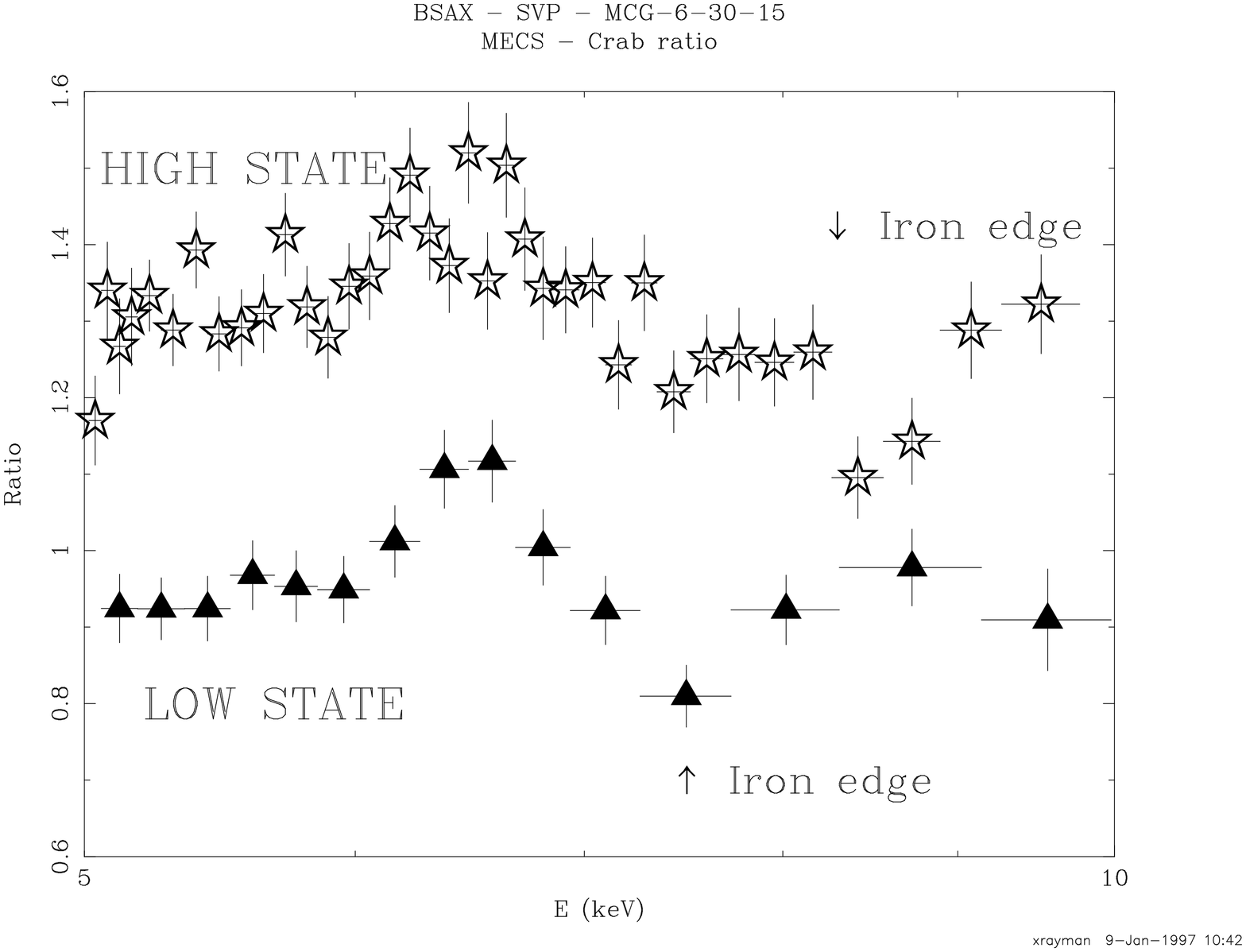,height=5.5truecm,width=8.8truecm,bbllx=25.0mm,bblly=13mm,bburx=201mm,bbury=275mm,angle=-90,clip=}
}
}
\caption{Ratio, in arbitrary units, of the MECS MCG 6-30-15 to Crab Spectrum.
The open stars represent the high state and the filled triangles the low state}
\end{figure}
A further complication comes from the Fe K-edge around 8 keV,
observed by GINGA and not seen by ASCA. Inspection of our spectrum shows some 
interesting but rather complex structure in the residuals around 8 keV. 
The picture simplifies considerably if we divide our 
observation in a low (source countrate lower than 1 cts/s) and a 
high state (source countrate higher than 1 cts/s) 
and accumulate separate spectra. 
As shown in Figure 3 the edge is present and variable in energy:
in the low state it is centered below 8 keV while in the high state 
above 8 keV. A plausible interpretation is that the ionization state
of the reprocessing matter which supposedly surrounds the primary source
varies as a function of the intensity of the continuum. The effect of an edge of 
variable energy when accumulating a spectrum of the entire observation would be to 
produce a depression in the continuum in the region between 7 and 9 keV. 

\subsection{ NGC 4151}

NGC 4151 has been observed for about 40 ks during the SVP. The observation 
is divided into 2 periods separated by about 200 ks. In the first 
period the source countrate for the 3 MECS is about 1.2 cts/s 
increasing to 2.1 cts/s in the second period. A description
of the broad band spectrum can be found in Piro et al. in these 
proceedings. In this presentation we focus on the MECS
data. In Figure 4 we report the low (first period) and high 
(second period) state spectra of NGC 4151. To avoid calibration 
uncertainties the reported spectra are obtained by dividing the NGC 4151 
spectra by the Crab spectrum. The shape of the
continuum  is different in the two states, in the sense that the source is 
softer when it is brighter, as seen in previous observations of
this source (e.g. Perola et al 1986, Yaqoob \& Warwick 1991). On the 
contrary the K$_\alpha$ line intensity
has not changed substantially (a similar behaviour  has been found in 
GINGA data Yaqoob \& Warwick 1991). This result indicates that the region 
responsible for the emission of the line does not experience a significant
variation of the ionizing continuum between the first and second period.
A possible interpretation is that the size of the line emitting region 
is larger then 200 kilo lightseconds.  

\begin{figure}
\centerline{
\hbox{
\psfig{figure=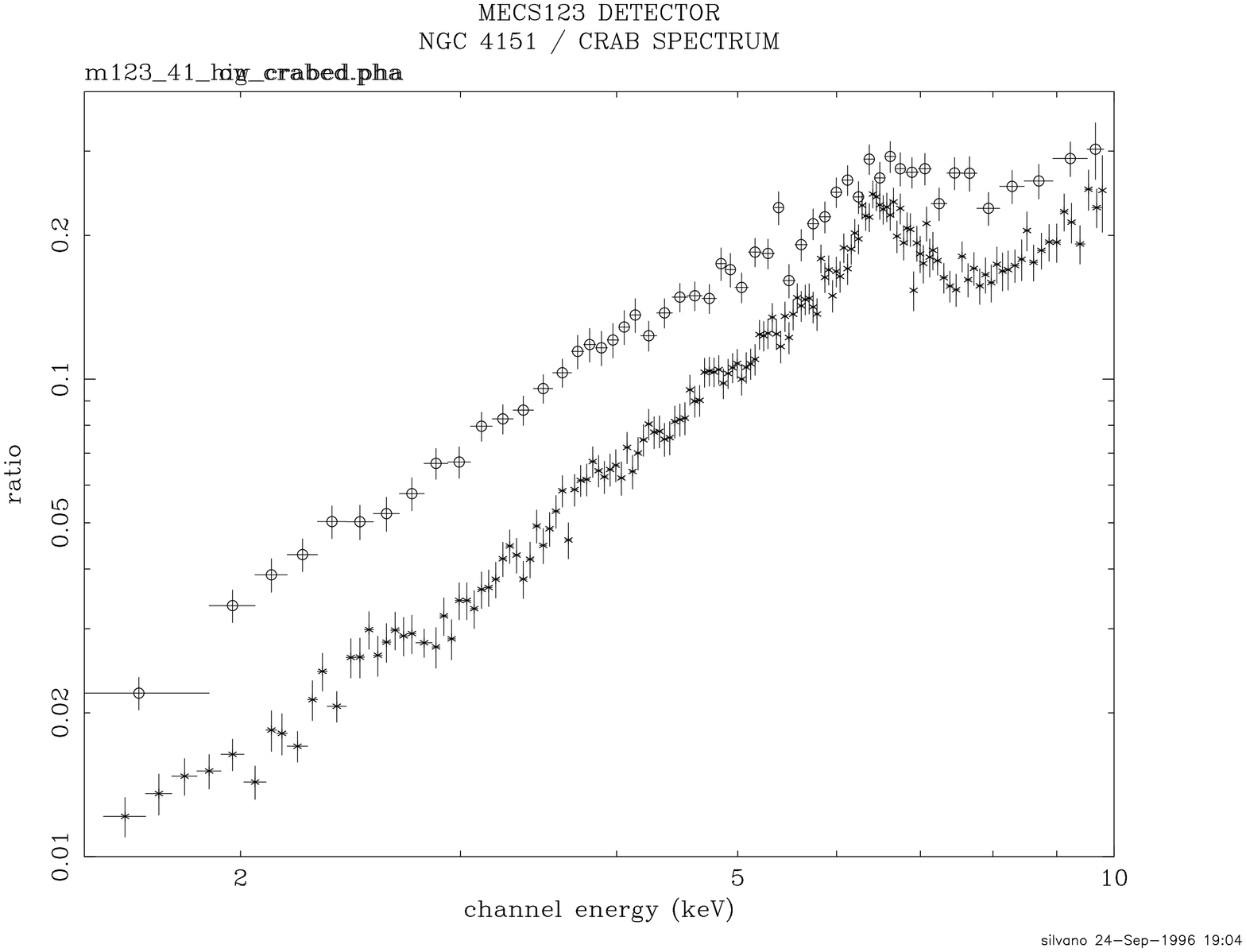,height=5.5truecm,width=8.8truecm,bbllx=27.5mm,bblly=13mm,bburx=201mm,bbury=275mm,angle=-90,clip=}
}
}
\caption{Ratio, in arbitrary units, of the NGC 4151 to Crab Spectrum.
The open circles represent the high state and the crosses the low state.}
\end{figure}

\subsection{The Cosmic X-ray Background}
While the satellite was recovering from a minor failure in July 1996, SAX was 
placed in default pointing mode, that is with the solar panels pointing 
directly towards the sun and the Narrow Field
Instruments (NFI) pointing in the direction of Polaris.
\begin{figure}
\centerline{
\hbox{
\psfig{figure=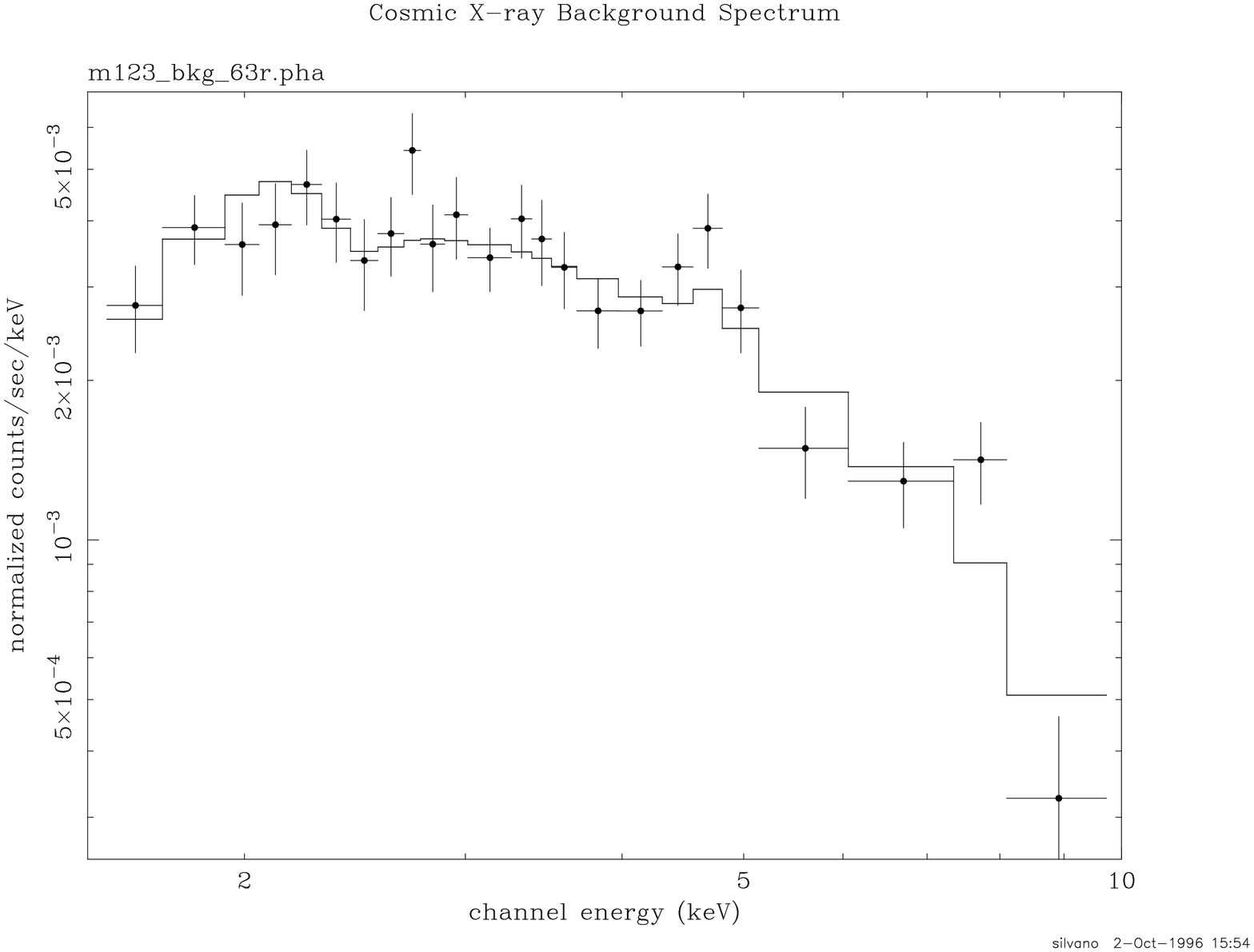,height=5.5truecm,width=8.8truecm,bbllx=27.5mm,bblly=13mm,bburx=201mm,bbury=275mm,angle=-90,clip=}
}
}
\caption{Spectrum of the Cosmic X-ray Background, the normalization is in arbitrary units.
 The solid line represents the best fitting absorbed power-law model.}
\end{figure}
During this period a long (100 ks) exposure was performed with the MECS.
Since Polaris is not a detectable X-ray source this observation can be 
considered as a first deep pointing of a "blank" area of the high galactic 
latitude sky. The limiting sensitivity of the field, roughly estimated as 
the flux of the weakest detectable source, is $\simeq$ $6\times 10^{-14}$
erg cm$^{-2}$s$^{-1}$ in the 2-10 keV band. A first inspection of the central 
and most 
sensitive region of the detector reveals that the number of detected sources
is consistent with that estimated by ASCA (Ogasaka et al. 1996 and Cagnoni 
et al. these proc.)

In Figure 4 we show the Cosmic X-ray Spectrum obtained by subtracting the 
"dark earth" spectrum (i.e. a 100 ks spectrum obtained by summing up 
observing periods during which the satellite NFIs were pointing in the 
direction of the dark earth) from the Polaris spectrum. Both spectra have 
been accumulated from the central circular region of the 3 detectors using an 
extraction radius of 8.4 arcminutes. 
The Cosmic X-ray Spectrum was fitted with a power-law plus galactic 
absorption model yielding the following best fitting parameters for the photon 
index and the normalization:
$\Gamma = 1.47^{+0.11}_{-0.15}$ and $ A = 11.9^{+1.7}_{-2.0}$
keV cm$^{-2}$s$^{-1}$sr$^{-1}$keV$^{-1}$,
were the quoted intervals are  68.3\% confidence intervals on
2 interesting parameters (i.e. $\Delta \chi^2 = 2.3$)
These numbers are in agreement with those derived from simultaneous 
fitting of ROSAT + ASCA background spectra (Chen et al. 1997)

\acknowledgements
SM thanks the SDC staff for the friendly and stimulating 
environment during the many visits.




\end{document}

%% file: newpsfig.tex
\def\PsfigVersion{1.10}
\def\setDriver{\DvipsDriver} 
\ifx\undefined\psfig\else \fi
\let\LaTeXAtSign=\@
\let\@=\relax
\edef\psfigRestoreAt{\catcode`\@=\number\catcode`@\relax}
\catcode`\@=11\relax
\newwrite\@unused
\def\ps@typeout#1{{\let\protect\string\immediate\write\@unused{#1}}}

\def\DvipsDriver{
	\ps@typeout{psfig/tex \PsfigVersion -dvips}
\def\PsfigSpecials{\DvipsSpecials} 	\def\ps@dir{/}
\def\ps@predir{} }
\def\OzTeXDriver{
	\ps@typeout{psfig/tex \PsfigVersion -oztex}
	\def\PsfigSpecials{\OzTeXSpecials}
	\def\ps@dir{:}
	\def\ps@predir{:}
	\catcode`\^^J=5
}

\def\figurepath{./:}

\def\DoPaths#1{\expandafter\EachPath#1\stoplist}
\def\leer{}
\def\EachPath#1:#2\stoplist{
  \ExistsFile{#1}{\SearchedFile}
  \ifx#2\leer
  \else
    \expandafter\EachPath#2\stoplist
  \fi}
\def\ps@dir{/}
\def\ExistsFile#1#2{%
   \openin1=\ps@predir#1\ps@dir#2
   \ifeof1
       \closein1
   \else
       \closein1
        \ifx\ps@founddir\leer
           \edef\ps@founddir{#1}
        \fi
   \fi}
\def\get@dir#1{%
  \def\ps@founddir{}
  \def\SearchedFile{#1}
  \DoPaths\figurepath
}

\def\@nnil{\@nil}
\def\@empty{}
\def\@psdonoop#1\@@#2#3{}
\def\@psdo#1:=#2\do#3{\edef\@psdotmp{#2}\ifx\@psdotmp\@empty \else
    \expandafter\@psdoloop#2,\@nil,\@nil\@@#1{#3}\fi}
\def\@psdoloop#1,#2,#3\@@#4#5{\def#4{#1}\ifx #4\@nnil \else
       #5\def#4{#2}\ifx #4\@nnil \else#5\@ipsdoloop #3\@@#4{#5}\fi\fi}
\def\@ipsdoloop#1,#2\@@#3#4{\def#3{#1}\ifx #3\@nnil 
       \let\@nextwhile=\@psdonoop \else
      #4\relax\let\@nextwhile=\@ipsdoloop\fi\@nextwhile#2\@@#3{#4}}
\def\@tpsdo#1:=#2\do#3{\xdef\@psdotmp{#2}\ifx\@psdotmp\@empty \else
    \@tpsdoloop#2\@nil\@nil\@@#1{#3}\fi}
\def\@tpsdoloop#1#2\@@#3#4{\def#3{#1}\ifx #3\@nnil 
       \let\@nextwhile=\@psdonoop \else
      #4\relax\let\@nextwhile=\@tpsdoloop\fi\@nextwhile#2\@@#3{#4}}
\ifx\undefined\fbox
\newdimen\fboxrule
\newdimen\fboxsep
\newdimen\ps@tempdima
\newbox\ps@tempboxa
\long\def\fbox#1{\leavevmode\setbox\ps@tempboxa\hbox{#1}\ps@tempdima\fboxrule
    \advance\ps@tempdima \fboxsep \advance\ps@tempdima \dp\ps@tempboxa
   \hbox{\lower \ps@tempdima\hbox
  {\vbox{\hrule height \fboxrule
          \hbox{\vrule width \fboxrule \hskip\fboxsep
          \vbox{\vskip\fboxsep \box\ps@tempboxa\vskip\fboxsep}\hskip 
                 \fboxsep\vrule width \fboxrule}
                 \hrule height \fboxrule}}}}
\fi
\newread\ps@stream
\newif\ifnot@eof       
\newif\if@noisy        
\newif\if@atend        
\newif\if@psfile       
{\catcode`\%=12\global\gdef\epsf@start{
\def\epsf@PS{PS}
\def\epsf@getbb#1{%
\openin\ps@stream=\ps@predir#1
\ifeof\ps@stream\ps@typeout{Error, File #1 not found}\else
   {\not@eoftrue \chardef\other=12
    \def\do##1{\catcode`##1=\other}\dospecials \catcode`\ =10
    \loop
       \if@psfile
	  \read\ps@stream to \epsf@fileline
       \else{
	  \obeyspaces
          \read\ps@stream to \epsf@tmp\global\let\epsf@fileline\epsf@tmp}
       \fi
       \ifeof\ps@stream\not@eoffalse\else
       \if@psfile\else
       \expandafter\epsf@test\epsf@fileline:. \\%
       \fi
          \expandafter\epsf@aux\epsf@fileline:. \\%
       \fi
   \ifnot@eof\repeat
   }\closein\ps@stream\fi}%
\long\def\epsf@test#1#2#3:#4\\{\def\epsf@testit{#1#2}
			\ifx\epsf@testit\epsf@start\else
\ps@typeout{Warning! File does not start with `\epsf@start'.  It may not be a PostScript file.}
			\fi
			\@psfiletrue} 
{\catcode`\%=12\global\let\epsf@percent=
\long\def\epsf@aux#1#2:#3\\{\ifx#1\epsf@percent
   \def\epsf@testit{#2}\ifx\epsf@testit\epsf@bblit
	\@atendfalse
        \epsf@atend #3 . \\%
	\if@atend	
	   \if@verbose{
		\ps@typeout{psfig: found `(atend)'; continuing search}
	   }\fi
        \else
        \epsf@grab #3 . . . \\%
        \not@eoffalse
        \global\no@bbfalse
        \fi
   \fi\fi}%
\def\epsf@grab #1 #2 #3 #4 #5\\{%
   \global\def\epsf@llx{#1}\ifx\epsf@llx\empty
      \epsf@grab #2 #3 #4 #5 .\\\else
   \global\def\epsf@lly{#2}%
   \global\def\epsf@urx{#3}\global\def\epsf@ury{#4}\fi}%
\def\epsf@atendlit{(atend)} 
\def\epsf@atend #1 #2 #3\\{%
   \def\epsf@tmp{#1}\ifx\epsf@tmp\empty
      \epsf@atend #2 #3 .\\\else
   \ifx\epsf@tmp\epsf@atendlit\@atendtrue\fi\fi}

\chardef\psletter = 11 
\chardef\other = 12

\newif \ifdebug 
\newif\ifc@mpute 
\c@mputetrue 

\let\then = \relax
\def\r@dian{pt }
\let\r@dians = \r@dian
\let\dimensionless@nit = \r@dian
\let\dimensionless@nits = \dimensionless@nit
\def\internal@nit{sp }
\let\internal@nits = \internal@nit
\newif\ifstillc@nverging
\def \Mess@ge #1{\ifdebug \then \message {#1} \fi}

{ 
	\catcode `\@ = \psletter
	\gdef \nodimen {\expandafter \n@dimen \the \dimen}
	\gdef \term #1 #2 #3%
	       {\edef \t@ {\the #1}
		\edef \t@@ {\expandafter \n@dimen \the #2\r@dian}%
		\t@rm {\t@} {\t@@} {#3}%
	       }
	\gdef \t@rm #1 #2 #3%
	       {{%
		\count 0 = 0
		\dimen 0 = 1 \dimensionless@nit
		\dimen 2 = #2\relax
		\Mess@ge {Calculating term #1 of \nodimen 2}%
		\loop
		\ifnum	\count 0 < #1
		\then	\advance \count 0 by 1
			\Mess@ge {Iteration \the \count 0 \space}%
			\Multiply \dimen 0 by {\dimen 2}%
			\Mess@ge {After multiplication, term = \nodimen 0}%
			\Divide \dimen 0 by {\count 0}%
			\Mess@ge {After division, term = \nodimen 0}%
		\repeat
		\Mess@ge {Final value for term #1 of 
				\nodimen 2 \space is \nodimen 0}%
		\xdef \Term {#3 = \nodimen 0 \r@dians}%
		\aftergroup \Term
	       }}
	\catcode `\p = \other
	\catcode `\t = \other
	\gdef \n@dimen #1pt{#1} 
}

\def \Divide #1by #2{\divide #1 by #2} 

\def \Multiply #1by #2
       {{
	\count 0 = #1\relax
	\count 2 = #2\relax
	\count 4 = 65536
	\Mess@ge {Before scaling, count 0 = \the \count 0 \space and
			count 2 = \the \count 2}%
	\ifnum	\count 0 > 32767 
	\then	\divide \count 0 by 4
		\divide \count 4 by 4
	\else	\ifnum	\count 0 < -32767
		\then	\divide \count 0 by 4
			\divide \count 4 by 4
		\else
		\fi
	\fi
	\ifnum	\count 2 > 32767 
	\then	\divide \count 2 by 4
		\divide \count 4 by 4
	\else	\ifnum	\count 2 < -32767
		\then	\divide \count 2 by 4
			\divide \count 4 by 4
		\else
		\fi
	\fi
	\multiply \count 0 by \count 2
	\divide \count 0 by \count 4
	\xdef \product {#1 = \the \count 0 \internal@nits}%
	\aftergroup \product
       }}

\def\r@duce{\ifdim\dimen0 > 90\r@dian \then   
		\multiply\dimen0 by -1
		\advance\dimen0 by 180\r@dian
		\r@duce
	    \else \ifdim\dimen0 < -90\r@dian \then  
		\advance\dimen0 by 360\r@dian
		\r@duce
		\fi
	    \fi}

\def\Sine#1%
       {{%
	\dimen 0 = #1 \r@dian
	\r@duce
	\ifdim\dimen0 = -90\r@dian \then
	   \dimen4 = -1\r@dian
	   \c@mputefalse
	\fi
	\ifdim\dimen0 = 90\r@dian \then
	   \dimen4 = 1\r@dian
	   \c@mputefalse
	\fi
	\ifdim\dimen0 = 0\r@dian \then
	   \dimen4 = 0\r@dian
	   \c@mputefalse
	\fi
	\ifc@mpute \then
		\divide\dimen0 by 180
		\dimen0=3.141592654\dimen0
		\dimen 2 = 3.1415926535897963\r@dian 
		\divide\dimen 2 by 2 
		\Mess@ge {Sin: calculating Sin of \nodimen 0}%
		\count 0 = 1 
		\dimen 2 = 1 \r@dian 
		\dimen 4 = 0 \r@dian 
		\loop
			\ifnum	\dimen 2 = 0 
			\then	\stillc@nvergingfalse 
			\else	\stillc@nvergingtrue
			\fi
			\ifstillc@nverging 
			\then	\term {\count 0} {\dimen 0} {\dimen 2}%
				\advance \count 0 by 2
				\count 2 = \count 0
				\divide \count 2 by 2
				\ifodd	\count 2 
				\then	\advance \dimen 4 by \dimen 2
				\else	\advance \dimen 4 by -\dimen 2
				\fi
		\repeat
	\fi		
			\xdef \sine {\nodimen 4}%
       }}

\def\Cosine#1{\ifx\sine\UnDefined\edef\Savesine{\relax}\else
		             \edef\Savesine{\sine}\fi
	{\dimen0=#1\r@dian\advance\dimen0 by 90\r@dian
	 \Sine{\nodimen 0}
	 \xdef\cosine{\sine}
	 \xdef\sine{\Savesine}}}	      

\def\psdraft{
	\def\@psdraft{0}
}
\def\psfull{
	\def\@psdraft{100}
}

\psfull

\newif\if@scalefirst
\def\psscalefirst{\@scalefirsttrue}
\def\psrotatefirst{\@scalefirstfalse}
\psrotatefirst

\newif\if@draftbox
\def\psnodraftbox{
	\@draftboxfalse
}
\def\psdraftbox{
	\@draftboxtrue
}
\@draftboxtrue

\newif\if@prologfile
\newif\if@postlogfile
\def\pssilent{
	\@noisyfalse
}
\def\psnoisy{
	\@noisytrue
}
\psnoisy
\newif\if@bbllx
\newif\if@bblly
\newif\if@bburx
\newif\if@bbury
\newif\if@height
\newif\if@width
\newif\if@rheight
\newif\if@rwidth
\newif\if@angle
\newif\if@clip
\newif\if@verbose
\def\@p@@sclip#1{\@cliptrue}
\newif\if@decmpr
\def\@p@@sfigure#1{\def\@p@sfile{null}\def\@p@sbbfile{null}\@decmprfalse
   \openin1=\ps@predir#1
   \ifeof1
	\closein1
	\get@dir{#1}
	\ifx\ps@founddir\leer
		\openin1=\ps@predir#1.bb
		\ifeof1
			\closein1
			\get@dir{#1.bb}
			\ifx\ps@founddir\leer
				\ps@typeout{Can't find #1 in \figurepath}
			\else
				\@decmprtrue
				\def\@p@sfile{\ps@founddir\ps@dir#1}
				\def\@p@sbbfile{\ps@founddir\ps@dir#1.bb}
			\fi
		\else
			\closein1
			\@decmprtrue
			\def\@p@sfile{#1}
			\def\@p@sbbfile{#1.bb}
		\fi
	\else
		\def\@p@sfile{\ps@founddir\ps@dir#1}
		\def\@p@sbbfile{\ps@founddir\ps@dir#1}
	\fi
   \else
	\closein1
	\def\@p@sfile{#1}
	\def\@p@sbbfile{#1}
   \fi
}
\def\@p@@sfile#1{\@p@@sfigure{#1}}
\def\@p@@sbbllx#1{
		\@bbllxtrue
		\dimen100=#1
		\edef\@p@sbbllx{\number\dimen100}
}
\def\@p@@sbblly#1{
		\@bbllytrue
		\dimen100=#1
		\edef\@p@sbblly{\number\dimen100}
}
\def\@p@@sbburx#1{
		\@bburxtrue
		\dimen100=#1
		\edef\@p@sbburx{\number\dimen100}
}
\def\@p@@sbbury#1{
		\@bburytrue
		\dimen100=#1
		\edef\@p@sbbury{\number\dimen100}
}
\def\@p@@sheight#1{
		\@heighttrue
		\dimen100=#1
   		\edef\@p@sheight{\number\dimen100}
}
\def\@p@@swidth#1{
		\@widthtrue
		\dimen100=#1
		\edef\@p@swidth{\number\dimen100}
}
\def\@p@@srheight#1{
		\@rheighttrue
		\dimen100=#1
		\edef\@p@srheight{\number\dimen100}
}
\def\@p@@srwidth#1{
		\@rwidthtrue
		\dimen100=#1
		\edef\@p@srwidth{\number\dimen100}
}
\def\@p@@sangle#1{
		\@angletrue
		\edef\@p@sangle{#1} 
}
\def\@p@@ssilent#1{ 
		\@verbosefalse
}
\def\@p@@sprolog#1{\@prologfiletrue\def\@prologfileval{#1}}
\def\@p@@spostlog#1{\@postlogfiletrue\def\@postlogfileval{#1}}
\def\@cs@name#1{\csname #1\endcsname}
\def\@setparms#1=#2,{\@cs@name{@p@@s#1}{#2}}
\def\ps@init@parms{
		\@bbllxfalse \@bbllyfalse
		\@bburxfalse \@bburyfalse
		\@heightfalse \@widthfalse
		\@rheightfalse \@rwidthfalse
		\def\@p@sbbllx{}\def\@p@sbblly{}
		\def\@p@sbburx{}\def\@p@sbbury{}
		\def\@p@sheight{}\def\@p@swidth{}
		\def\@p@srheight{}\def\@p@srwidth{}
		\def\@p@sangle{0}
		\def\@p@sfile{} \def\@p@sbbfile{}
		\def\@p@scost{10}
		\def\@sc{}
		\@prologfilefalse
		\@postlogfilefalse
		\@clipfalse
		\if@noisy
			\@verbosetrue
		\else
			\@verbosefalse
		\fi
}
\def\parse@ps@parms#1{
	 	\@psdo\@psfiga:=#1\do
		   {\expandafter\@setparms\@psfiga,}}
\newif\ifno@bb
\def\bb@missing{
	\if@verbose{
		\ps@typeout{psfig: searching \@p@sbbfile \space  for bounding box}
	}\fi
	\no@bbtrue
	\epsf@getbb{\@p@sbbfile}
        \ifno@bb \else \bb@cull\epsf@llx\epsf@lly\epsf@urx\epsf@ury\fi
}	
\def\bb@cull#1#2#3#4{
	\dimen100=#1 bp\edef\@p@sbbllx{\number\dimen100}
	\dimen100=#2 bp\edef\@p@sbblly{\number\dimen100}
	\dimen100=#3 bp\edef\@p@sbburx{\number\dimen100}
	\dimen100=#4 bp\edef\@p@sbbury{\number\dimen100}
	\no@bbfalse
}
\newdimen\p@intvaluex
\newdimen\p@intvaluey
\def\rotate@#1#2{{\dimen0=#1 sp\dimen1=#2 sp
		  \global\p@intvaluex=\cosine\dimen0
		  \dimen3=\sine\dimen1
		  \global\advance\p@intvaluex by -\dimen3
		  \global\p@intvaluey=\sine\dimen0
		  \dimen3=\cosine\dimen1
		  \global\advance\p@intvaluey by \dimen3
		  }}
\def\compute@bb{
		\no@bbfalse
		\if@bbllx \else \no@bbtrue \fi
		\if@bblly \else \no@bbtrue \fi
		\if@bburx \else \no@bbtrue \fi
		\if@bbury \else \no@bbtrue \fi
		\ifno@bb \bb@missing \fi
		\ifno@bb \ps@typeout{FATAL ERROR: no bb supplied or found}
			\no-bb-error
		\fi
		\count203=\@p@sbburx
		\count204=\@p@sbbury
		\advance\count203 by -\@p@sbbllx
		\advance\count204 by -\@p@sbblly
		\edef\ps@bbw{\number\count203}
		\edef\ps@bbh{\number\count204}
		\if@angle 
			\Sine{\@p@sangle}\Cosine{\@p@sangle}
	        	{\dimen100=\maxdimen\xdef\r@p@sbbllx{\number\dimen100}
					    \xdef\r@p@sbblly{\number\dimen100}
			                    \xdef\r@p@sbburx{-\number\dimen100}
					    \xdef\r@p@sbbury{-\number\dimen100}}
                        \def\minmaxtest{
			   \ifnum\number\p@intvaluex<\r@p@sbbllx
			      \xdef\r@p@sbbllx{\number\p@intvaluex}\fi
			   \ifnum\number\p@intvaluex>\r@p@sbburx
			      \xdef\r@p@sbburx{\number\p@intvaluex}\fi
			   \ifnum\number\p@intvaluey<\r@p@sbblly
			      \xdef\r@p@sbblly{\number\p@intvaluey}\fi
			   \ifnum\number\p@intvaluey>\r@p@sbbury
			      \xdef\r@p@sbbury{\number\p@intvaluey}\fi
			   }
			\rotate@{\@p@sbbllx}{\@p@sbblly}
			\minmaxtest
			\rotate@{\@p@sbbllx}{\@p@sbbury}
			\minmaxtest
			\rotate@{\@p@sbburx}{\@p@sbblly}
			\minmaxtest
			\rotate@{\@p@sbburx}{\@p@sbbury}
			\minmaxtest
			\edef\@p@sbbllx{\r@p@sbbllx}\edef\@p@sbblly{\r@p@sbblly}
			\edef\@p@sbburx{\r@p@sbburx}\edef\@p@sbbury{\r@p@sbbury}
		\fi
		\count203=\@p@sbburx
		\count204=\@p@sbbury
		\advance\count203 by -\@p@sbbllx
		\advance\count204 by -\@p@sbblly
		\edef\@bbw{\number\count203}
		\edef\@bbh{\number\count204}
}
\def\in@hundreds#1#2#3{\count240=#2 \count241=#3
		     \count100=\count240	
		     \divide\count100 by \count241
		     \count101=\count100
		     \multiply\count101 by \count241
		     \advance\count240 by -\count101
		     \multiply\count240 by 10
		     \count101=\count240	
		     \divide\count101 by \count241
		     \count102=\count101
		     \multiply\count102 by \count241
		     \advance\count240 by -\count102
		     \multiply\count240 by 10
		     \count102=\count240	
		     \divide\count102 by \count241
		     \count200=#1\count205=0
		     \count201=\count200
			\multiply\count201 by \count100
		 	\advance\count205 by \count201
		     \count201=\count200
			\divide\count201 by 10
			\multiply\count201 by \count101
			\advance\count205 by \count201
		     \count201=\count200
			\divide\count201 by 100
			\multiply\count201 by \count102
			\advance\count205 by \count201
		     \edef\@result{\number\count205}
}
\def\compute@wfromh{
		\in@hundreds{\@p@sheight}{\@bbw}{\@bbh}
		\edef\@p@swidth{\@result}
}
\def\compute@hfromw{
	        \in@hundreds{\@p@swidth}{\@bbh}{\@bbw}
		\edef\@p@sheight{\@result}
}
\def\compute@handw{
		\if@height 
			\if@width
			\else
				\compute@wfromh
			\fi
		\else 
			\if@width
				\compute@hfromw
			\else
				\edef\@p@sheight{\@bbh}
				\edef\@p@swidth{\@bbw}
			\fi
		\fi
}
\def\compute@resv{
		\if@rheight \else \edef\@p@srheight{\@p@sheight} \fi
		\if@rwidth \else \edef\@p@srwidth{\@p@swidth} \fi
}
\def\compute@sizes{
	\compute@bb
	\if@scalefirst\if@angle
	\if@width
	   \in@hundreds{\@p@swidth}{\@bbw}{\ps@bbw}
	   \edef\@p@swidth{\@result}
	\fi
	\if@height
	   \in@hundreds{\@p@sheight}{\@bbh}{\ps@bbh}
	   \edef\@p@sheight{\@result}
	\fi
	\fi\fi
	\compute@handw
	\compute@resv}
\def\OzTeXSpecials{
	\special{empty.ps /@isp {true} def}
	\special{empty.ps \@p@swidth \space \@p@sheight \space
			\@p@sbbllx \space \@p@sbblly \space
			\@p@sbburx \space \@p@sbbury \space
			startTexFig \space }
	\if@clip{
		\if@verbose{
			\ps@typeout{(clip)}
		}\fi
		\special{empty.ps doclip \space }
	}\fi
	\if@angle{
		\if@verbose{
			\ps@typeout{(rotate)}
		}\fi
		\special {empty.ps \@p@sangle \space rotate \space} 
	}\fi
	\if@prologfile
	    \special{\@prologfileval \space } \fi
	\if@decmpr{
		\if@verbose{
			\ps@typeout{psfig: Compression not available
			in OzTeX version \space }
		}\fi
	}\else{
		\if@verbose{
			\ps@typeout{psfig: including \@p@sfile \space }
		}\fi
		\special{epsf=\ps@predir\@p@sfile \space }
	}\fi
	\if@postlogfile
	    \special{\@postlogfileval \space } \fi
	\special{empty.ps /@isp {false} def}
}
\def\DvipsSpecials{
	\special{ps::[begin] 	\@p@swidth \space \@p@sheight \space
			\@p@sbbllx \space \@p@sbblly \space
			\@p@sbburx \space \@p@sbbury \space
			startTexFig \space }
	\if@clip{
		\if@verbose{
			\ps@typeout{(clip)}
		}\fi
		\special{ps:: doclip \space }
	}\fi
	\if@angle
		\if@verbose{
			\ps@typeout{(clip)}
		}\fi
		\special {ps:: \@p@sangle \space rotate \space} 
	\fi
	\if@prologfile
	    \special{ps: plotfile \@prologfileval \space } \fi
	\if@decmpr{
		\if@verbose{
			\ps@typeout{psfig: including \@p@sfile.Z \space }
		}\fi
		\special{ps: plotfile "`zcat \@p@sfile.Z" \space }
	}\else{
		\if@verbose{
			\ps@typeout{psfig: including \@p@sfile \space }
		}\fi
		\special{ps: plotfile \@p@sfile \space }
	}\fi
	\if@postlogfile
	    \special{ps: plotfile \@postlogfileval \space } \fi
	\special{ps::[end] endTexFig \space }
}
\def\psfig#1{\vbox {
	\ps@init@parms
	\parse@ps@parms{#1}
	\compute@sizes
	\ifnum\@p@scost<\@psdraft{
		\PsfigSpecials 
		\vbox to \@p@srheight sp{
			\hbox to \@p@srwidth sp{
				\hss
			}
		\vss
		}
	}\else{
		\if@draftbox{		
			\hbox{\fbox{\vbox to \@p@srheight sp{
			\vss
			\hbox to \@p@srwidth sp{ \hss 
			 \hss }
			\vss
			}}}
		}\else{
			\vbox to \@p@srheight sp{
			\vss
			\hbox to \@p@srwidth sp{\hss}
			\vss
			}
		}\fi

	}\fi
}}
\psfigRestoreAt
\setDriver
\let\@=\LaTeXAtSign